\documentclass{article}

\usepackage{arxiv}
\usepackage[utf8]{inputenc} 
\usepackage[T1]{fontenc}    
\usepackage{hyperref}       
\usepackage{url}            
\usepackage{booktabs}       
\usepackage{amsfonts}       
\usepackage{nicefrac}       
\usepackage{microtype}      
\usepackage{lipsum}
\usepackage{graphicx}
\usepackage{subcaption}
\usepackage{float}

\title{Assessing the effects of exposure to sulfuric acid aerosol on respiratory function in adults}

\author{
  Lamin ~Juwara \\
  Department of Epidemiology, Biostatistics \\ and Occupational Health \\
  McGill University\\
  Montreal, QC, Canada  \\
  \texttt{lamin.juwara@mail.mcgill.ca} \\
   \And
 Jennifer ~Boateng  \\
  Department of Neurology \& Neurosurgery  \\
  McGill University\\
  Montreal, QC, Canada \\
  \texttt{Jennifer.boateng@mail.mcgill.ca} \\
}

\begin{document}
\maketitle

\begin{abstract}
\textbf{Background:} Sulfuric acid aerosol is suspected to be a major contributor to mortality and morbidity associated with air pollution.

\textbf{Objective:} To determine if exposure of human participants to anticipated levels of sulfuric acid aerosol ($\sim 100 \mu g/m^3$) in the near future would have an adverse effect on respiratory function.

\textbf{Methods:} We used data from 28 adults exposed to sulfuric acid for 4 hours in a controlled exposure chamber over a 3 day period with repeated measures of pulmonary function (FEV1) recorded at 2 hour intervals. Measurements were also recorded after 2 and 24 hours post exposure. We formulated a linear mixed effect model for FEV1 with fixed effects (day of treatment, hour, day-hour interaction, and smoking status), a random intercept and an AR1 covariance structure to estimate the effect of aerosol exposure on FEV1. We further assessed whether smoking status modified the exposure effects and compared the analysis to the method used by Kerr et al.,1981. 

\textbf{Results:} The effect of day 3 exposure  is negatively associated with lung function (coefficient ($\beta$), -0.08; 95\% CI, -0.16 to -0.01). A weak negative association is observed with increasing hours of exposure ($\beta$, -0.01; 95\% CI, -0.03 to 0.00). Among the smokers, we found a significant negative association with hours of exposure ($\beta$, -0.02; 95\% CI, -0.03 to -0.00), day 3 exposure ($\beta$, -0.11; 95\% CI, -0.14 to -0.02) and a borderline adverse effect for day 2 treatment ($\beta$, -0.06; 95\% CI, -0.14 to 0.03) whilst no significant association was observed for nonsmokers. 

\textbf{Conclusions:} Anticipated deposits of sulfuric acid aerosol in the near would adversely affect respiratory function. The effect observed in smokers is significantly more adverse than in nonsmokers. 
\end{abstract}


\section{Introduction}

Sulfuric acid ($\texttt{H}_2\texttt{SO}_4$) is considered a possible cause of increased mortality and morbidity resulting from various episodes of air pollutions over the past.\cite{lawther1963compliance}\cite{chan1997characterisation} The recent increases in $\texttt{H}_2\texttt{SO}_4$ emission from automobile catalytic converters present enormous environmental challenges with peak estimates projected to reach as high as $80\mu g/m^3$  in some industrialized countries. \cite{kerr1993evidence} 

The effect of sulfuric acid aerosol on pulmonary function in humans has been demonstrated in various studies over the past. \cite{kerr1981effects} In controlled studies \cite{amdur1958mechanics}\cite{amdur1952toxicity} where participants were exposed to high levels of $\texttt{H}_2\texttt{SO}_4$ ($\ge 200$ micro grams per meter-cube, $\mu g/m^3$), adverse effects in pulmonary functions were associated with $\texttt{H}_2\texttt{SO}_4$. However, studies of the association of lung function and low levels of $\texttt{H}_2\texttt{SO}_4$ have generally reported inconsistent findings. Kerr et al. (1981), in their analyses, reported no significant difference in pulmonary function in a randomized study where participants were exposed to 100 $\mu g/m^3$ of $\texttt{H}_2\texttt{SO}_4$ in an environmentally controlled chamber. 

The objectives of the current study include (1) To determine if exposure (of participants) to anticipated levels of $\texttt{H}_2\texttt{SO}_4$ aerosol in the near future, in a realistic time frame, would have an adverse effect upon respiratory function; (2) To assess if estimated effects of $\texttt{H}_2\texttt{SO}_4$ exposure are modified by smoking status; (3) To compare the results obtained to the analyses conducted in Kerr et al., 1981. 
	
\section{Methods}
\subsection{Exposure and Outcome definition}
The study design, experimental set-up and methods used are described in (Kerr et al., 1981). Additionally, information on other pulmonary function measurements not included in the current report are provided in \cite{kerr1981effects}.

\underline{Subject selection}

The study was based on 28 healthy adults (aged 18- 45 years) with no previous history of chronic respiratory or cardiovascular diseases. The participants comprise of 19 males and 9 females, 14 smokers and 14 non-smokers of mean age 24 and mean height 175m. The participants were required to refrain from smoking the morning prior to the study.

\underline{Air Pollution Exposure}

The study was conducted in an environmentally controlled exposure chamber. Over a 3-day period and at the same time, the participants were exposed to: (1) treatment day 1, the subjects breathed in only filtered clean air for 6 hours; 2) treatment day 2, they breathed in $100 \mu g/m^3$ of sulfuric acid for 4 hours; and 3) treatment day 3, the subjects breathed in only filtered clean air for 6 hours. At the time of study, the participants were blinded to the type of exposure administered in any day.

\underline{Outcome Measures }

Measurement of pulmonary test function (FEV1) by spirometry was performed prior to exposure, at 2 hours during exposure, immediately following exposure (approximately 4 hours from the starts), and 2 hours post exposure. The same measurements were repeated for each participant for three consecutive days. To simulate an environment similar to living in an urban setting, at 1 hour and 3 hours during exposure in each day, the participants were required to complete a bicycle ergometer exercise using a load of 100W at 60rpm for 15 minutes.

\underline{Additional covariates }

Information was also collected on other relevant covariates which were taken into account during study design; they include age, height, sex, day of treatment, time of measurements and smoking status. All the variables included in the study had no missing measurement.

\subsection{Statistical Analysis }

Statistical analyses were performed using R Version 3.4.2 (R Foundation for Statistical Computing, Vienna, Austria). In this analysis, we formulated a linear mixed effect (LME) model to fit the repeated FEV1 measure. To properly specify the LME model, we used the four-stage approach recommended by Diggle and Verbeke, 2002 and others.\cite{diggle2002analysis}\cite{molenberghs2000model}.

\begin{figure}[h]
		\caption{ FEV1 repeated measures for (n=28) study participants for treatment days 1,2,3. Time point 0 represent measurement at baseline (prior to exposure), 1-4 denotes FEV1 levels for hours 1,2,3 and 4 respectively, 5 represents FEV1 level at 2 hours after treatment and time point 6 is 24 hours after treatment.}
	\label{fig:repeatedmeasures}
	\begin{subfigure}[a]{0.31\textwidth}
		\includegraphics[width=\textwidth,height=5cm]{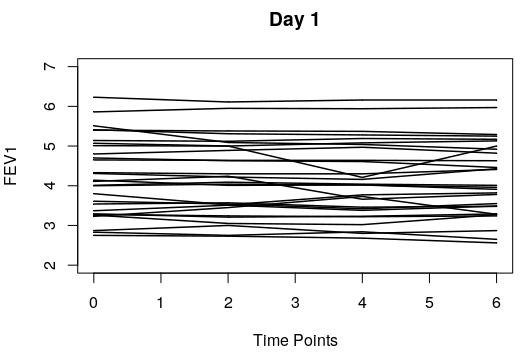}
	\end{subfigure}
	~ %
	\begin{subfigure}[a]{0.31\textwidth}
		\includegraphics[width=\textwidth,height=5cm]{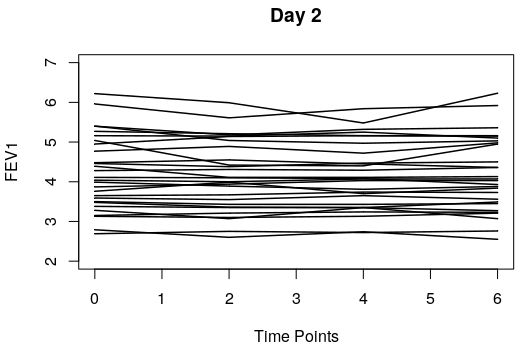}
	\end{subfigure}
	~ %
\begin{subfigure}[a]{0.31\textwidth}
	\includegraphics[width=\textwidth,height=5cm]{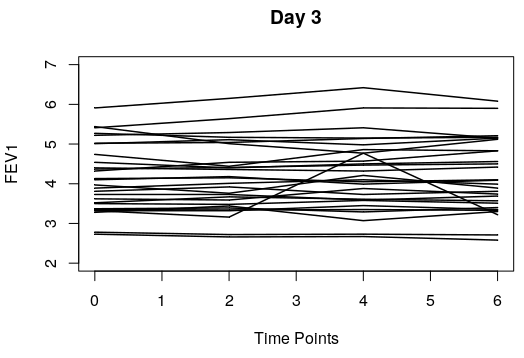}
\end{subfigure}
\subcaption*{\tiny{ FEV1 = measure of pulmonary test function; Day $i$ =  treatment day $i$ for  $i\in\{1,2,3\}$}}
\end{figure}

\bigskip

\begin{figure}[h]
\caption{ Mean FEV1 measures for treatment days 1, 2 and 3.}
	\label{fig:meanmeasures}
	\centering
	\includegraphics[width=0.5\linewidth, height=0.22\textheight]{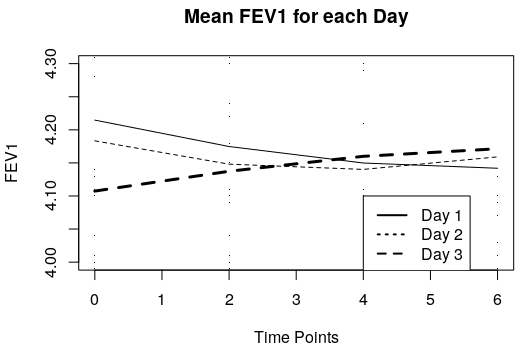}
\end{figure}

In the exploratory analysis, we plot repeated FEV1 measures of  participants in each of the three treatment groups. In Figure  \ref{fig:repeatedmeasures}, we present individual FEV1 measurements for treatment day 1 (left), day 2 (middle) and day 3 (right).  The day curves for measures of FEV1 appear to be linear with time suggesting they could be included in the model as a linear covariate. In Figure \ref{fig:meanmeasures}, we present an exploration of the mean FEV1 curves for each treatment day over the time course of treatment. The mean curves suggest linearity in mean FEV1 over treatment duration. In the LME model, we proposed a mean model component involving linear fixed effects of the treatment groups (day), time (hour), treatment-time interaction (day-hour interaction), and smoking status. 

\begin{table}[h]
	\centering
\caption{Covariance and empirical correlation estimates for FEV1 repeated measures data. Covariance above the diagonal, variance on the diagonal and correlation below the diagonal.}
\label{fig:covariancematrix}
\includegraphics[width=0.4\linewidth]{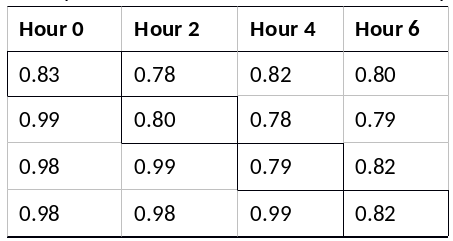}
\end{table}

The second stage involves specifying a covariance structure to proper capture variations between individual participants and the covariance between FEV1 measurements at different times on the same participant. Residuals obtained from a linear fit of the mean component specified in the first stage are used to construct an empirical correlation matrix of the 6-hour time-period. The estimates of the variance between participants in the same treatment group (day) are printed in the diagonals of the covariance matrix in Table \ref{fig:covariancematrix}. 

\begin{figure}[h]
	\centering
		\caption{Scatter plots of FEV1 repeated measures at measurements time points 0 hours (baseline), 2 hours, 4 hours and 6 hours (2 hours post exposure).}
	\label{fig:scatterfev}
	\includegraphics[width=0.7\linewidth]{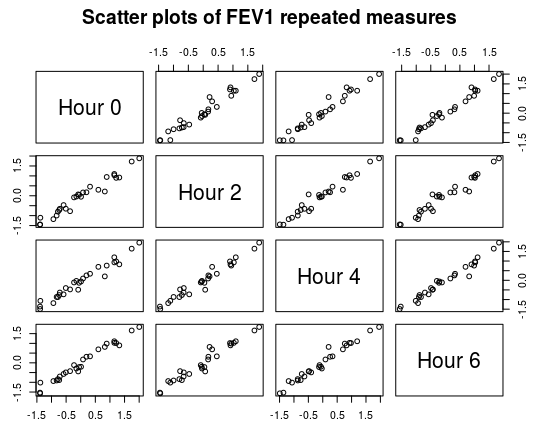}
\end{figure}

\bigskip

The patterns observed from the exploratory plots and examination of the correlation structure seem to suggest a model with random intercepts with measurements taken close apart being more correlated. Additionally, a scatter plot of FEV1 measures at different times in Figure \ref{fig:scatterfev} suggest a homogeneous variability between individuals. Consequently, we propose an autoregressive order 1 (AR1) covariance structure and random intercepts term for the participants. 

We combine the mean model component and variance structure proposed in stage I-II to formulate a linear mixed model for  FEV1 of the form
\begin{equation}
\texttt{FEV1}_{ijk} = \beta_0 + \beta_1\texttt{Smoker} + \alpha_i\texttt{Day} + \gamma_k\texttt{Hour} + \mu_{ik}\texttt{Day}\ast\texttt{Hour} + b_{ij} + \epsilon_{ijk}   
\end{equation} 

where   $\beta_0$ represents a common constant for all measurements,  $\beta_1$ is the coefficient of smoking status, $\alpha_i$ is the parameter for treatment day $i$, $\gamma_k$ is the parameter corresponding to the hour $k$ and $\mu_{ik}$ is the coefficient of the interaction between day $i$ and hour $k$. We assume the random intercept $b_{ij}$  is normally distributed with mean zero and constant variance $\sigma_b$, the measurement error $\epsilon_{ijk}$ is normally distributed with mean zero and variance $\sigma_\epsilon$ and independent of $b_{ij}$. In the third and fourth stages, we fit the mean model and explore the possibility of incorporating polynomial day curves over treatment hours.

To assess the presence of potential effect modification resulting from the smoking status of the participant, we separately analyze FEV1 measurements for smokers and nonsmokers, and assess the difference in exposure effect between smokers and nonsmokers. 

Finally, we examine model diagnostic plots to access the LME model assumptions. The normality assumption of the random intercept and measurement error are assessed from Q-Q plots of the random intercept in Figure \ref{fig:diagnostics}. The plots of standardized residuals are also presented in Figure 6 to assess the assumption of linearity. Further, the appropriateness of the specified autocorrelation is also assessed via examination of ACF plots of the normalized residuals of the fitted model. In order to ensure that the specified correlation structure is indeed ideal, the fitted model is compared with a model incorporation a compound symmetric covariance structure (a potential candidate for the covariance structure) and the model with the best fit is consequently selected by based on AIC criterion.

\section{Results}

Participant characteristics in the three treatment days are described in the methods section. Out of the 28 participants, 14 smokers and 14 nonsmokers, the mean FV1 measurement on treatment day 1 varied from a maximum of 4.23 prior to filtered air exposure to 4.15 after 4 hours of exposure. A similar varying mean FEV1 measure is observed on day 3, under the same filtered air treatment, with a maximum mean measure of 4.23 observed immediately post exposure and the least measure just before exposure. On day 2, when the participants were exposed to 100 of sulfuric acid, we observe a low mean FEV1 measure  of 4.14 after two and four hours of exposure; the highest mean FEV1 measure was 4.22 on exposure day 2 is observed. More specifically, we observe a decreasing trend in pulmonary lung function in the day 2 mean plot in Figure 2; the trend begins to improve with improving FEV1 measure post exposures. 
In Table 1, we present the covariance and correlation of mean FEV1 measures between the different measurement times.  The between-patient variances within day group at each measurement time are seen in the diagonals of the covariance matrix ranging from 0.79-0.83 and the correlation below the diagonals range from 0.98-1.  In general, we observe a decreasing empirical correlation estimate from 0.99 between FEV1 at time Hour 0 and Hour 2 to 0.98 between FEV1 at time Hour 0 and HR 6.  A Scatter plot of FEV1 repeated measures at each 2-hour versus FEV1 all other measurement times are presented in Figure 3.
Based on results obtain from the exploratory data analysis, we proposed linear mixed model FEV1 in equation (1). The model included smoking status, day of treatment, hour (time), day-hour interaction, and a random intercept with an AR1 covariance structure.  Results obtained from a fit of the linear mixed model proposed in equation one is presented in Table \ref{fig:lme}.

\begin{table}[H]
	\centering
	\caption{Linear mixed effect model for FEV1}
	\label{fig:lme}
	\includegraphics[width=0.6\linewidth]{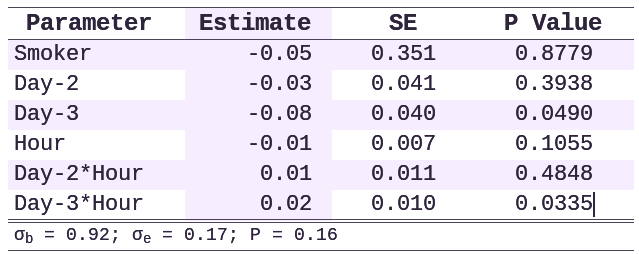}
\end{table}

Exposure to sulfuric acid in treatment day 2 is seemingly weakly associated with a decrease in FEV1 measures with an effect estimate (95\% CI) of -0.03(-0.12, 0.04). Compared to the baseline day 1 exposure, we see a significant adverse association between exposure to filtered air on day 3 and pulmonary function measure with an effect estimate (95\% CI) of -0.08(-0.16,-0.01). Further, with increasing time, we see a seemingly negative association between exposure to sulfuric acid and pulmonary function over the course of the study with a somewhat weaker significant effect estimate -0.01(-0.03,0.00). We see similar weak, but positive, associations between FEV1 and the effect of interaction between sulfuric acid exposure on day 2 and  hour 0.01(-0.01,0.03) or the interaction between day 3 and hour estimate 0.02(0.00,0.04). 

An assessment of potential effect modification by smoking status indicates a significant evidence of modification of the association between FEV1 measure and treatment day 3, and the association of FEV1 with time (hour). Results obtained from separately analyzing data for smokers and nonsmokers is in Table \ref{fig:lmesmoker}. As expected, in the group of participants who smoke, we observe a significant adverse association between FEV1 measures and time of exposure with an estimated effect (95\% CI) -0.02(-0.03,-0.00). A similar negative association of FEV1 measure and exposure to filtered clean air on day 3 is observed (coefficient -0.11, 95\% CI -0.14 to -0.02).

\begin{table}[H]
	\centering
	\caption{Linear mixed effect model estimates for FEV1 independently assessed for smokers and nonsmokers}
	\label{fig:lmesmoker}
	\includegraphics[width=0.8\linewidth]{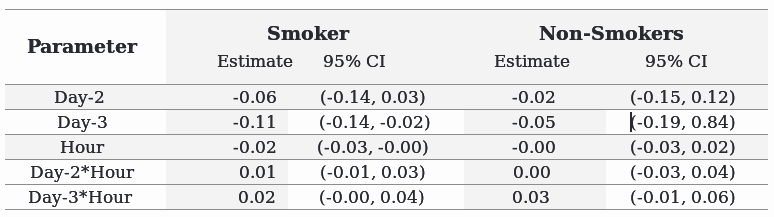}
\end{table}

\begin{figure}[H]
	\caption{Autocorrelation plot of LME model. Normalized residual (right) and non-normalized (left)}
	\label{fig:autocorr-plot}
	\centering
	\includegraphics[width=0.5\linewidth,height=5cm]{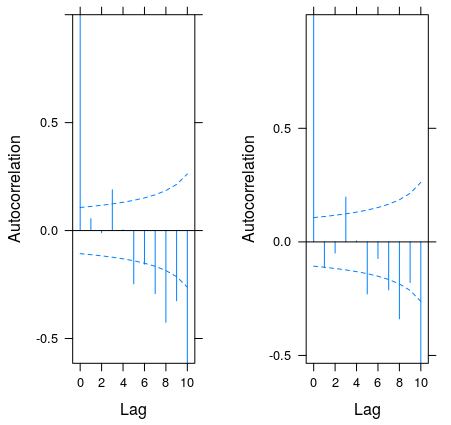}
\end{figure}

\begin{figure}[H]
	\caption{ Normal Q-Q plots of random intercept (left); Q-Q plots of treatment days 1-3 (middle); Residual plot for treatment days (right) }
	\label{fig:diagnostics}
	\centering
	\begin{subfigure}[a]{0.31\textwidth}
		\includegraphics[width=\textwidth,height=5cm]{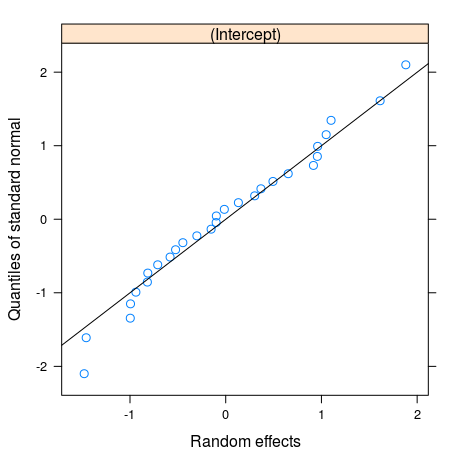}
	\end{subfigure}
	~ %
	\begin{subfigure}[a]{0.31\textwidth}
		\includegraphics[width=\textwidth,height=5cm]{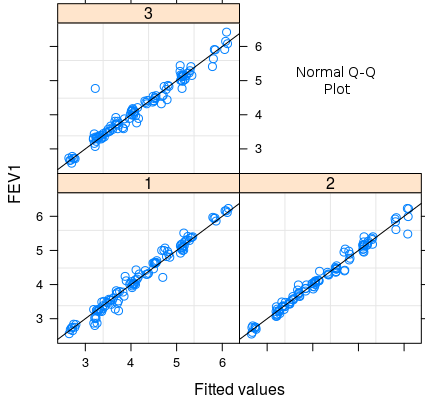}
	\end{subfigure}
	~ %
	\begin{subfigure}[a]{0.31\textwidth}
		\includegraphics[width=\textwidth,height=5cm]{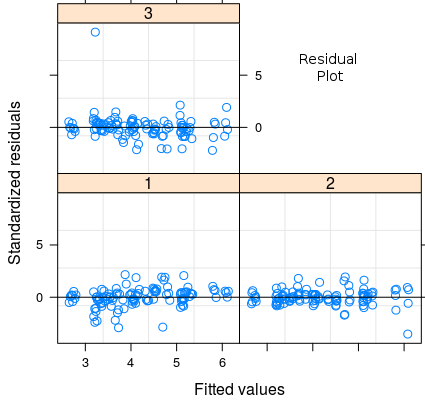}
	\end{subfigure}
\end{figure}

In Figure \ref{fig:autocorr-plot}, we show a diagnostic plot of the assessment of the AR1 correlation structure. The plots of the normalized residuals indicate that our model has dealt with temporal autocorrelation at lag 1 and 2. However, we observe some unexpected values for larger lags probably resulting from other anomalies in the data. The correlation pattern was further assessed by examining a variogram plot of the residuals. Additionally, residual plots of the model and random intercept, in Figure \ref{fig:diagnostics} \ (right), do not give any indication that the normality assumptions are violated. Moreover, a plot of the observed and fitted values in Figure \ref{fig:diagnostics} \ indicates a suitable fit of the model to data.

\section{Discussion}

Results obtained in this analysis indicate that short-term exposure to low levels of sulfuric acid aerosol has an adverse effect on the measure of pulmonary lung function (FEV1). More specifically, we see a significant negative association between FEV1 and treatment with filtered clean on the last day of the trial. A similar seemingly negative, but weak, association is observed with time (hour) of exposure to sulfuric acid. A stronger adverse effect is evident in participants who smoke compared to nonsmokers.

Previous studies \cite{lee2014air}\cite{franchini2009particulate} have suggested that sulfuric acid constitutes the main component of air pollution responsible for increased mortality and morbidity in various events of major air pollution over the recent past. Spektor et al., 1989 \cite{spektor1989effect} showed that cumulative exposure of inhaled sulfuric acid on is adversely associated with tracheobronchial particle clearance in healthy humans. Similar studies \cite{amdur1952toxicity}\cite{amdur1958mechanics}\cite{lee2014air} over the years support the general consensus that exposure to high levels of sulfuric acid could lead to complications with pulmonary system. However, results from \cite{kerr1981effects} looking at exposure to low doses of sulfuric acid exposure have been inconsistent.
The analysis in this report explored the possibility of effect modification by smoking status of the participant. As expected, participants with previous smoking history appear more adversely affected by exposure to sulfuric acid compared to nonsmokers. This is indicative of a major flaw in the design of the study, which inadvertently affected the analysis in Kerr et al. (1981). In our analysis, results obtained from fitting a model for smokers among the participants were significantly different from the results obtained for nonsmokers. In general, we observe a weak adverse effect of exposure to sulfuric acid on pulmonary lung functions. These findings are in line with previous work using higher concentrations of the exposure primarily in animals. 

In the analysis of the dataset presented by Kerr at al., 1981, they reported no significant difference in pulmonary function during exposure, immediately after, or 2 and 24 hours post exposure. In their analysis of the  data, they used a paired student t-test to compare the exposed participant with their controls. They assumed that the design was completely randomized and balanced. Indeed, the analysis would have been suitable if the measurements were not repeated and adjustments were only needed for within-subject variations. Consequently, a limitation of the method they used is that it does not account for repeated measurements within a subject and potential between-subject variations. They also used ANOVA for factorial design to assess the varying effect of pulmonary lung function with day of exposure, hour and day-hour interaction; however, the method is often criticized to be biased in general and loses information by not effectively incorporating intermediate measurements. Moreover, it could lead to inaccurate conclusions in unbalanced designs.\cite{gurka2011avoiding} \ As a consequence of the patterns of correlations observed in the exploratory analysis, a standard analysis of variance as prescribed in Milliken and Johnson\cite{kerr1993evidence} is likely not appropriate for this dataset. Thus, a linear mixed effect analysis implemented, which did not in general assume a complete balance setting. A major advantage of using a linear mixed model for this design is that the manner in which the subjects are assigned to the treatment days in itself typically induces a covariance structure. Moreover, the design induces a covariance due to the contributions of the random effects. In the analysis, the 28 participants were included in the analysis and any potential confounding resulting from smoking was controlled by including their smoking status in the model.

In order to accurately capture the true effect of exposure to sulfuric acid on pulmonary lung function, information on other potentially relevant variables would need to be properly included in the mean model.  Incorporating fundamental covariates such as age, sex and BMI could immensely improve the model fit. The accuracy of the present analysis, in part, assumes that during the design stage as reported in Kerr et al. (1980), the effect of the variables age, height and sex were controlled. Another possible limitation of the model is in the specification of the covariance structure. Although the model with the best covariance structure was selected (AR1 VS Compound symmetric), the number of repeated measures was too small to truly capture and account for the covariance structure of the data generating mechanism.\cite{cheng2010real}

We conclude that short-term exposure to sulfuric acid is negatively associated with the measure of pulmonary lung function FEV1; this is particularly evident among smokers. Since the study was conducted over a short period of time using a low dosage of sulfuric acid, we observe a minimally weak adverse effect in nonsmokers. This seems to be in agreement with studies \cite{spektor1989effect}\cite{wolff1986effects} that assessed the association between FEV1 and exposure to various sulfuric constituents of air pollutants. A further investigation incorporating other potentially useful risk factor or longer exposure time is warranted.

\bibliographystyle{unsrt}  
\bibliography{references}  

\begin{thebibliography}{10}

\bibitem{lawther1963compliance}
PJ~Lawther.
\newblock Compliance with the clean air act: medical aspects.
\newblock {\em J. Inst. Fuel}, 36:341, 1963.

\bibitem{chan1997characterisation}
YC~Chan, RW~Simpson, GH~McTainsh, PD~Vowles, DD~Cohen, and GM~Bailey.
\newblock Characterisation of chemical species in pm2. 5 and pm10 aerosols in
  brisbane, australia.
\newblock {\em Atmospheric Environment}, 31(22):3773--3785, 1997.

\bibitem{kerr1993evidence}
JB~Kerr and CT~McElroy.
\newblock Evidence for large upward trends of ultraviolet-b radiation linked to
  ozone depletion.
\newblock {\em Science}, 262(5136):1032--1034, 1993.

\bibitem{kerr1981effects}
HD~Kerr, TJ~Kulle, BP~Farrell, LR~Sauder, JL~Young, DL~Swift, and RM~Borushok.
\newblock Effects of sulfuric acid aerosol on pulmonary function in human
  subjects: an environmental chamber study.
\newblock {\em Environmental research}, 26(1):42--50, 1981.

\bibitem{amdur1958mechanics}
Mary~O Amdur and Jere Mead.
\newblock Mechanics of respiration in unanesthetized guinea pigs.
\newblock {\em American Journal of Physiology-Legacy Content}, 192(2):364--368,
  1958.

\bibitem{amdur1952toxicity}
Mary~O Amdur, RZ~Schulz, Ph~Drinker, et~al.
\newblock Toxicity of sulfuric acid mist to guinea pigs.
\newblock {\em Arch. Indust. Hyg. \& Occupational Med.}, 5(4):318--29, 1952.

\bibitem{diggle2002analysis}
Peter Diggle, Peter~J Diggle, Patrick Heagerty, Patrick~J Heagerty, Kung-Yee
  Liang, Scott Zeger, et~al.
\newblock {\em Analysis of longitudinal data}.
\newblock Oxford University Press, 2002.

\bibitem{molenberghs2000model}
Geert Molenberghs and Geert Verbeke.
\newblock A model for longitudinal data.
\newblock {\em Linear Mixed Models for Longitudinal Data}, pages 19--29, 2000.

\bibitem{lee2014air}
BJ~Lee, B~Kim, and K~Lee.
\newblock Air pollution exposure and cardiovascular disease, toxicol. res, 30
  (2), p71-75, 2014.

\bibitem{franchini2009particulate}
Massimo Franchini and Pier~Mannuccio Mannucci.
\newblock Particulate air pollution and cardiovascular risk: short-term and
  long-term effects.
\newblock In {\em Seminars in thrombosis and hemostasis}, volume~35, pages
  665--670. {\copyright} Thieme Medical Publishers, 2009.

\bibitem{spektor1989effect}
Dalia~M Spektor, Bai~M Yen, and Morton Lippmann.
\newblock Effect of concentration and cumulative exposure of inhaled sulfuric
  acid on tracheobronchial particle clearance in healthy humans.
\newblock {\em Environmental health perspectives}, 79:167--172, 1989.

\bibitem{gurka2011avoiding}
Matthew~J Gurka, Lloyd~J Edwards, and Keith~E Muller.
\newblock Avoiding bias in mixed model inference for fixed effects.
\newblock {\em Statistics in medicine}, 30(22):2696--2707, 2011.

\bibitem{cheng2010real}
Jing Cheng, Lloyd~J Edwards, Mildred~M Maldonado-Molina, Kelli~A Komro, and
  Keith~E Muller.
\newblock Real longitudinal data analysis for real people: building a good
  enough mixed model.
\newblock {\em Statistics in medicine}, 29(4):504--520, 2010.

\bibitem{wolff1986effects}
RK~Wolff, RF~Henderson, RH~Gray, RL~Carpenter, and FF~Hahn.
\newblock Effects of sulfuric acid mist inhalation on mucous clearance and on
  airway fluids of rats and guinea pigs.
\newblock {\em Journal of Toxicology and Environmental Health, Part A Current
  Issues}, 17(1):129--142, 1986.

\end{thebibliography}


\end{document}